\documentclass[twocolumn,showpacs,preprintnumbers,aps,prd,floatfix]{revtex4}
\usepackage{graphicx}
\usepackage{dcolumn}
\usepackage{amsfonts}
\usepackage{latexsym}
\usepackage{amssymb} 
\usepackage{verbatim}
\usepackage{amsthm}
\usepackage{amsmath}
\def\apj #1 #2 #3 {#1, ApJ, {\bf #2}, #3}
\def\apjl #1 #2 #3 {#1, ApJ, {\bf #2}, L#3}
\def\apjs #1 #2 #3 {#1, ApJS, {\bf #2}, #3}
\def\aap  #1 #2 #3 {#1, A\&A, {\bf #2}, #3}
\def\mnras #1 #2 #3 {#1, MNRAS, {\bf #2}, #3}
\def\pra #1 #2 #3 {#1, Phys.~Rev.~A., {\bf #2}, #3}
\def\prb #1 #2 #3 {#1, Phys.~Rev.~B., {\bf #2}, #3}
\def\prc #1 #2 #3 {#1, Phys.~Rev.~C., {\bf #2}, #3}
\def\prd #1 #2 #3 {#1, Phys.~Rev.~D., {\bf #2}, #3}
\def\pre #1 #2 #3 {#1, Phys.~Rev.~E., {\bf #2}, #3}
\def\prl #1 #2 #3 {#1, Phys.~Rev.~Lett., {\bf #2}, #3}
\def\plb #1 #2 #3 {#1, Phys.~Lett.~B., {\bf #2}, #3}
\def\science #1 #2 #3 {#1, Science., {\bf #2}, #3}
\def\nature #1 #2 #3 {#1, Nature., {\bf #2}, #3}
\def\nphysa #1 #2 #3 {#1, Nucl.~Phys.~A., {\bf #2}, #3}
\def\nphysb #1 #2 #3 {#1, Nucl.~Phys.~B., {\bf #2}, #3}
\def\nphysbs #1 #2 #3 {#1, Nucl.~Phys.~B.~Suppl., {\bf #2}, #3}

\def\h#1{\hbox{${}^{#1}$H}}

\def\h502{\hbox{$ h^{2}_{50}$}}

\def\fun#1#2{\lower3.6pt\vbox{\baselineskip0pt\lineskip.9pt
  \ialign{$\mathsurround=0pt#1\hfil##\hfil$\crcr#2\crcr\sim\crcr}}}
\newcommand{\MR}{{{\mathbb R}}}

\begin{document}
%
\title{
  Gauge-invariant Formulation of the Second-order Cosmological Perturbations
} 
\author{Kouji Nakamura   \footnote{E-mail address: kouchan@th.nao.ac.jp}}
\affiliation{%
Department of Astronomical Science, the Graduate University for
Advanced Studies, 2-21-1, Osawa, Mitaka, Tokyo 181-8588, Japan}
\date{\today}
\begin{abstract}
Gauge invariant treatments of the second order cosmological
perturbation in a four dimensional homogeneous isotropic
universe filled with the perfect fluid are completely formulated
without any gauge fixing.
We derive all components of the Einstein equations in the case
where the first order vector and tensor modes are negligible.
These equations imply that the tensor and the vector mode of the
second order metric perturbations may be generated by the
scalar-scalar mode coupling of the linear order perturbations as
the result of the non-linear effects of the Einstein equations. 
\end{abstract}
\maketitle


To clarify the relation between scenarios of the early universe
and observational data such as the cosmic microwave background
(CMB) anisotropies, the general relativistic cosmological 
{\it linear} perturbation theory has been developed to a high
degree of sophistication during the last 25
years\cite{Bardeen-1980}.
Recently, the first order approximation of the early universe
from a homogeneous isotropic one is revealed by the observation
of the CMB by Wilkinson Microwave Anisotropy Probe
(WMAP)\cite{WMAP} and is suggested that fluctuations in the
early universe are adiabatic and Gaussian at least in the first
order approximation.
One of the next theoretical tasks is to clarify the accuracy of
these results, for example, through the non-Gaussianity.
To accomplish this, the {\it second order} cosmological
perturbation theory is necessary.
So, the perturbation theory beyond the linear order has been
investigated by many
authors\cite{Tomita-1967,M.Bruni-S.Soonego-CQG1997,S.Sonego-M.Bruni-CMP1998,other-branch}
and is a topical subject, in particular, to study the
non-Gaussianity generated during the
inflation\cite{Non-Gaussianity-inflation} and that will be
observed in CMB data\cite{Non-Gaussianity-in-CMB}.


In this letter, we show the gauge invariant formulation of the
general relativistic second order cosmological perturbations on
Friedmann-Robertson-Walker (FRW) universe filled with the
perfect fluid.
This short letter is prepared to show the aspect of the
special importance of our companion
paper\cite{kouchan-cosmology-second}, briefly. 
The formulation in this paper is one of the applications of the
gauge invariant formulation of the second order perturbation
theory on a generic background spacetime developed in
Refs.\cite{kouchan-gauge-inv,kouchan-cosmology-second}.
We treat all perturbative variables in the gauge invariant
manner.
We also derive all components of the second order Einstein
equations of cosmological perturbations in terms of these gauge
invariant variables without any gauge fixing.


First, we explain the ``{\it gauge}'' in general relativistic
perturbations\cite{J.M.Stewart-M.Walker11974,M.Bruni-S.Soonego-CQG1997,S.Sonego-M.Bruni-CMP1998}. 
To explain this, we have to explain what we are doing in
perturbation theories, at first. 
In perturbation theories, we treat two spacetimes.
One is the physical spacetime ${\cal M}$ which we will describe
by perturbations.
In cosmological perturbations, ${\cal M}$ is an universe with
small inhomogeneities.
Another is the background spacetime ${\cal M}_{0}$ which is
prepared for perturbative analyses.
In cosmological perturbations, ${\cal M}_{0}$ is the FRW
universe with the metric
\begin{eqnarray}
  g_{ab} = a^{2}(\eta)\left(
    - (d\eta)_{a}(d\eta)_{b}
    + \gamma_{ij}(dx^{i})_{a}(dx^{j})_{b}
  \right),
  \label{eq:background-metric}
\end{eqnarray}
where $\gamma_{ij}$ is the metric on a maximally symmetric
3-space with curvature constant $K$.
We note that ${\cal M}$ and ${\cal M}_{0}$ are different
manifolds.


We also note that, in perturbation theories, we always write
equations in the form 
\begin{equation}
  \label{eq:variable-symbolic-perturbation}
  Q(``p\mbox{''}) = Q_{0}(p) + \delta Q(p).
\end{equation}
Eq.~(\ref{eq:variable-symbolic-perturbation}) gives the relation
between variables on ${\cal M}$ and ${\cal M}_{0}$.
Actually, $Q(``p\mbox{''})$ in the right hand side (rhs) of
Eq.~(\ref{eq:variable-symbolic-perturbation}) is a variable on
${\cal M}$, while $Q_{0}(p)$ and $\delta Q(p)$ in the left hand
side (lhs) of Eq.~(\ref{eq:variable-symbolic-perturbation}) are
those on ${\cal M}_{0}$.
Further, the point $``p\mbox{''}$ in $Q(``p\mbox{''})$ is on
${\cal M}$, while the point $p$ in $Q_{0}(p)$ and $\delta Q(p)$
is on ${\cal M}_{0}$.
Since Eq.~(\ref{eq:variable-symbolic-perturbation}) is a field
equation, we implicitly identify these two points $``p\mbox{''}$
and $p$ through Eq.~(\ref{eq:variable-symbolic-perturbation}) by
a map 
${\cal M}_{0}\rightarrow{\cal M}$ $:$ $p\in{\cal M}_{0}\mapsto
``p\mbox{''}\in{\cal M}$.
This identification is the ``{\it gauge choice}'' in
perturbation theories\cite{J.M.Stewart-M.Walker11974}.


Moreover, the above gauge choice between ${\cal M}_{0}$ and
${\cal M}$ is not unique in theories with general
covariance.
Rather, Eq.~(\ref{eq:variable-symbolic-perturbation}) involves
the degree of freedom corresponding to the choice of the map
${\cal X}$ $:$ ${\cal M}_{0}\mapsto{\cal M}$.
This degree of freedom is the ``{\it gauge degree of freedom}''
in general relativistic perturbation theory.
By virture of the general covariance, there is no preferred coordinate
system and we have no guiding principle to choose the
identification map ${\cal X}$.
So, we may always change the gauge choice ${\cal X}$.


To develop this understanding of the ``gauge'', we introduce a
parameter $\lambda$ for the perturbation and the 4+1-dimensional
manifold ${\cal N}={\cal M}\times\MR$ so that 
${\cal M}_{0}=\left.{\cal N}\right|_{\lambda=0}$ and 
${\cal M}={\cal M}_{\lambda}=\left.{\cal N}\right|_{\MR=\lambda}$.
On this extended manifold ${\cal N}$, the gauge choice is
regarded as a diffeomorphism ${\cal X}_{\lambda}$ $:$ ${\cal N}$
$\rightarrow$ ${\cal N}$ such that ${\cal X}_{\lambda}$ $:$
${\cal M}_{0}$ $\rightarrow$ ${\cal M}_{\lambda}$.
Further, a gauge choice ${\cal X}_{\lambda}$ is introduced as an
exponential map with the generator ${}^{{\cal X}}\!\eta^{a}$
which is chosen so that its integral curve in ${\cal N}$ is
transverse to each ${\cal M}_{\lambda}$ everywhere on 
${\cal N}$\cite{M.Bruni-S.Soonego-CQG1997}.
Points lying on the same integral curve are regarded as the
``{\it same point}'' by the gauge choice ${\cal X}_{\lambda}$.


Now, we define the first and the second order perturbation of
the physical variable $Q$ on ${\cal M}_{\lambda}$ by the
pulled-back variable ${\cal X}_{\lambda}^{*}Q$ on
${\cal M}_{0}$, which is induced by the gauge choice
${\cal X}_{\lambda}$ and expanded as
\begin{eqnarray}
  {\cal X}^{*}_{\lambda}Q_{\lambda}
  =
  Q_{0}
  + \lambda \left.{\pounds}_{{}^{{\cal X}}\!\eta}Q\right|_{{\cal M}_{0}}
  + \frac{1}{2} \lambda^{2} 
  \left.{\pounds}_{{}^{{\cal X}}\!\eta}^{2}Q\right|_{{\cal M}_{0}}
  + O(\lambda^{3}).
  \label{eq:Taylor-expansion-of-calX}
\end{eqnarray}
$Q_{0}=\left.Q\right|_{{\cal M}_{0}}$ is the background value of
$Q$ and all terms in Eq.~(\ref{eq:Taylor-expansion-of-calX}) are
evaluated on ${\cal M}_{0}$.
Since Eq.~(\ref{eq:Taylor-expansion-of-calX}) is just the
perturbative expansion of ${\cal X}^{*}_{\lambda}Q_{\lambda}$,
we may regard that the first and the second order perturbations
of $Q$ are given by 
${}^{(1)}_{\;\cal X}\!Q=\left.{\pounds}_{{}^{\cal X}\!\eta}Q\right|_{{\cal M}_{0}}$,
and 
${}^{(2)}_{\;\cal X}\!Q=\left.{\pounds}_{{}^{\cal X}\!\eta}^{2}Q\right|_{{\cal M}_{0}}$,
respectively.


Suppose that we have two gauge choices ${\cal X}_{\lambda}$ and 
${\cal Y}_{\lambda}$ with the generators ${}^{\cal X}\!\eta^{a}$
and ${}^{\cal Y}\!\eta^{a}$, respectively.
When these generators have the different tangential
components to each ${\cal M}_{\lambda}$, these 
${\cal X}_{\lambda}$ and ${\cal Y}_{\lambda}$ are regarded as
``{\it different gauge choices}''.
Further, the ``{\it gauge transformation}'' is regarded as the
change of the gauge choice 
${\cal X}_{\lambda}\rightarrow{\cal Y}_{\lambda}$, which is
given by the diffeomorphism
\begin{equation}
  \label{eq:diffeo-def-from-Xinv-Y}
  \Phi_{\lambda} :=
  ({\cal X}_{\lambda})^{-1}\circ{\cal Y}_{\lambda}
  : {\cal M}_{0}\rightarrow {\cal M}_{0}
\end{equation}
The diffeomorphism $\Phi_{\lambda}$ does change the point
identification.
Further, $\Phi_{\lambda}$ induces a pull-back from the
representation ${}^{\cal X}Q_{\lambda}$ in the gauge choice
${\cal X}_{\lambda}$ to the representation 
${}^{\cal Y}Q_{\lambda}$ in the gauge choice 
${\cal Y}_{\lambda}$ as   
\begin{eqnarray}
  {}^{\cal Y}Q_{\lambda}
  =
  \left.{\cal Y}^{*}_{\lambda}Q\right|_{{\cal M}_{0}}
  =
  \left.
    \left(
      {\cal X}^{-1}_{\lambda}
      {\cal Y}_{\lambda}
    \right)^{*}
    \left(
      {\cal X}^{*}_{\lambda}Q
    \right)
  \right|_{{\cal M}_{0}}
  =  \Phi^{*}_{\lambda} {}^{\cal X}Q_{\lambda}.
  \label{eq:Bruni-45-one}
\end{eqnarray}
As discussed in Ref.\cite{S.Sonego-M.Bruni-CMP1998},
${}^{\cal Y}Q_{\lambda}=\Phi^{*}_{\lambda}{}^{\cal X}Q_{\lambda}$ 
is expanded as  
\begin{eqnarray}
  {}^{\cal Y}Q_{\lambda} &=& {}^{\cal X}\!Q
  + \lambda {\pounds}_{\xi_{(1)}} {}^{\cal X}\!Q
  \nonumber\\
  && \quad
  + \frac{\lambda^{2}}{2} \left(
    {\pounds}_{\xi_{(2)}} + {\pounds}_{\xi_{(1)}}^{2}
  \right) {}^{\cal X}\!Q
  + O(\lambda^{3}),
  \label{eq:Bruni-46-one} 
\end{eqnarray}
where $\xi_{(1)}^{a}$ and $\xi_{(2)}^{a}$ are the generators of
$\Phi_{\lambda}$.


Comparing Eq.~(\ref{eq:Bruni-46-one}) with 
${\cal Y}^{*}_{\lambda}\circ\left({\cal X}_{\lambda}^{-1}\right)^{*}{}^{{\cal X}}\!Q$
in terms of the generators ${}^{\cal X}\eta^{a}$ and
${}^{\cal Y}\eta^{a}$, we find 
$\xi_{(1)}^{a}={}^{\cal Y}\eta^{a}-{}^{\cal X}\eta^{a}$
and $\xi_{(2)}^{a}=\left[{}^{\cal Y}\eta,{}^{\cal X}\eta\right]^{a}$.
Further, Eqs.~(\ref{eq:Bruni-46-one}) and
(\ref{eq:Taylor-expansion-of-calX}) yield
\begin{eqnarray}
  \label{eq:Bruni-47-one}
  {}^{(1)}_{\;{\cal Y}}\!Q - {}^{(1)}_{\;{\cal X}}\!Q &=& 
  {\pounds}_{\xi_{(1)}}Q_{0}, \\
  \label{eq:Bruni-49-one}
  {}^{(2)}_{\;\cal Y}\!Q - {}^{(2)}_{\;\cal X}\!Q &=& 
  2 {\pounds}_{\xi_{(1)}} {}^{(1)}_{\;\cal X}\!Q 
  +\left\{{\pounds}_{\xi_{(2)}}+{\pounds}_{\xi_{(1)}}^{2}\right\} Q_{0}.
\end{eqnarray}
These are the gauge transformation rules for the first and the
second order perturbations, respectively.


Inspecting gauge transformation rules (\ref{eq:Bruni-47-one})
and (\ref{eq:Bruni-49-one}), we define the gauge invariant
variables of each order.
First, we consider the metric perturbations. 
The metric on ${\cal M}_{\lambda}$ is expanded through the gauge
choice ${\cal X}_{\lambda}$ as  
\begin{eqnarray}
  {\cal X}^{*}_{\lambda}\bar{g}_{ab}
  &=&
  g_{ab} + \lambda {}_{{\cal X}}\!h_{ab} 
  + \frac{\lambda^{2}}{2} {}_{{\cal X}}\!l_{ab}
  + O^{3}(\lambda).
  \label{eq:metric-expansion}
\end{eqnarray}
Since the $\eta=const.$ hypersurfaces in the metric
(\ref{eq:background-metric}) are maximally symmetric, the
components $\{h_{\eta\eta},h_{i\eta},h_{ij}\}$ 
of the first order metric perturbation $h_{ab}$ is classified 
into the three sets of variables 
$\{h_{\eta\eta},h_{(VL)},h_{(L)},h_{TL}\}$,
$\{{h_{(V)}}_{i},{h_{(TV)}}_{i}\}$, and
${h_{(TT)ij}}$, which are defined by 
\begin{eqnarray}
  h_{\eta i} &=:& D_{i}h_{(VL)} + h_{(V)i}, \quad D^{i}h_{(V)i} = 0
  , \nonumber\\
  h_{ij} &=:& a^{2} h_{(L)} \gamma_{ij} + a^{2}h_{(T)ij}, 
  \quad \gamma^{ij}{h_{(T)}}_{ij} = 0
  , \nonumber\\ 
  h_{(T)ij} &=:& 
  \left(D_{i}D_{j} - \frac{1}{3}\gamma_{ij}\Delta\right)h_{(TL)}
  \nonumber\\
  && \quad + 2 D_{(i}h_{(TV)j)} + {h_{(TT)ij}}
  \label{eq:hij-decomp}
  , \\
  && D^{i} h_{(TV)i} = 0, \quad D^{i} h_{(TT)ij} = 0.
  \nonumber
\end{eqnarray}
The uniqueness of the decomposition (\ref{eq:hij-decomp})
is guaranteed by the existence of the Green functions of
operators $\Delta:=D^{i}D_{i}$, $\Delta+2K$, and $\Delta+3K$,
where $D_{i}$ is the covariant derivative associated with the
metric $\gamma_{ij}$.
In terms of the variables given in (\ref{eq:hij-decomp}), we
define a vector field $X_{a}$ by 
\begin{eqnarray}
  \label{eq:Xa-def}
  X_{a} &:=& X_{\eta}(d\eta)_{a} + X_{i}(dx^{i})_{a}, \\
  X_{\eta} &:=& h_{(VL)} - \frac{1}{2} a^{2}\partial_{\tau}h_{(TL)} 
  ,
  \label{eq:Xeta-def}
  , \\
  X_{i} &:=&  
  a^{2} \left(
      h_{(TV)i}
    + \frac{1}{2} D_{i}h_{(TL)}
  \right)
  ,
  \label{eq:Xi-def}
\end{eqnarray}
where $X_{a}$ is transformed as
${}_{\;{\cal Y}}\!X_{a}-{}_{\;{\cal X}}\!X_{a} = \xi_{(1)a}$
under the gauge transformation (\ref{eq:Bruni-47-one}).
We also define the variables 
\begin{eqnarray}
  - 2 a^{2} \stackrel{(1)}{\Phi} 
  &:=& 
  h_{\eta\eta} - 2 \left( \partial_{\eta} - {\cal H} \right) X_{\eta}
  , \\
  - 2 a^{2} \stackrel{(1)}{\Psi}
  &:=& 
  a^{2} \left( h_{(L)} - \frac{1}{3}\Delta h_{(TL)} \right)
  + 2 {\cal H} X_{\eta}
  , \\
  a^{2} \stackrel{(1)\;\;}{\nu_{i}}
  &:=& h_{(V)i} - a^{2}\partial_{\eta}h_{(TV)i}
  , \\
  \stackrel{(1)\;\;\;}{\chi_{ij}} &:=&  h_{(TT)ij}
  ,
\end{eqnarray}
where ${\cal H} = \partial_{\eta}a/a$.
Further, we denote 
\begin{eqnarray}
  \label{eq:first-order-gauge-inv-metrc-pert-components}
  {\cal H}_{ab}
  &=& 
  - 2 a^{2} \stackrel{(1)}{\Phi} (d\eta)_{a}(d\eta)_{b}
  + 2 a^{2} \stackrel{(1)}{\nu_{i}} (d\eta)_{(a}(dx^{i})_{b)}
  \nonumber\\
  && \quad
  + a^{2} 
  \left( - 2 \stackrel{(1)}{\Psi} \gamma_{ij} 
    + \stackrel{(1)}{{\chi}_{ij}} \right)
  (dx^{i})_{a}(dx^{j})_{b},
\end{eqnarray}
where
$D^{i}\stackrel{(1)}{\nu_{i}}=\stackrel{(1)}{\chi_{[ij]}}=\stackrel{(1)}{\chi^{i}_{\;i}}=D^{i}\stackrel{(1)}{\chi_{ij}}=0$.
The tensor field ${\cal H}_{ab}$ is gauge invariant under the
gauge transformation (\ref{eq:Bruni-47-one}).
In the cosmological perturbations\cite{Bardeen-1980},
$\{\stackrel{(1)}{\Phi},\stackrel{(1)}{\Psi}\}$,
$\stackrel{(1)}{\nu_{i}}$, $\stackrel{(1)}{\chi_{ij}}$ are
called the scalar, vector, and tensor modes, respectively.
In terms of the variables ${\cal H}_{ab}$ and $X_{a}$, the
original first order metric perturbation $h_{ab}$ is given by 
\begin{eqnarray}
  h_{ab} =: {\cal H}_{ab} + {\pounds}_{X}g_{ab}.
  \label{eq:linear-metric-decomp}
\end{eqnarray}
Since the scalar mode dominates in the early universe, we assume
that $\stackrel{(1)}{\nu_{i}}=\stackrel{(1)}{\chi_{ij}}=0$ in
this letter.


Now, we consider the decomposition of the second order metric
perturbation $l_{ab}$ into the gauge invariant and variant
parts.
Through the above variables $X_{a}$ and $h_{ab}$, we first
consider the variable $\hat{L}_{ab}$ defined by 
\begin{eqnarray}
  \hat{L}_{ab}
  :=
  l_{ab} - 2 {\pounds}_{X}h_{ab} + {\pounds}_{X}^{2}g_{ab}
\end{eqnarray}
Under the gauge transformation rules (\ref{eq:Bruni-47-one}) and 
(\ref{eq:Bruni-49-one}), the variable $\hat{L}_{ab}$ is
transformed as
\begin{eqnarray}
  {}_{\;\cal Y}\!\hat{L}_{ab} - {}_{\;\cal X}\!\hat{L}_{ab}
  = 
  {\pounds}_{\sigma} g_{ab} ,
  \quad 
  \sigma^{a} := \xi_{(2)}^{a} + [\xi_{(1)},X]^{a}.
  \label{eq:4.67}
\end{eqnarray}
We note that this gauge transformation (\ref{eq:4.67}) have
the same form as that of $h_{ab}$.
Then, as shown above, there exist a vector field $Y_{a}$ and a
tensor field ${\cal L}_{ab}$ such that the second order metric
perturbation $l_{ab}$ is given by 
\begin{eqnarray}
  \label{eq:H-ab-in-gauge-X-def-second-1}
  l_{ab}
  =:
  {\cal L}_{ab} + 2 {\pounds}_{X} h_{ab}
  + \left(
      {\pounds}_{Y}
    - {\pounds}_{X}^{2} 
  \right)
  g_{ab}.
\end{eqnarray}
The variables ${\cal L}_{ab}$ and $Y^{a}$ are the gauge
invariant and variant parts of $l_{ab}$, respectively, and the
components of ${\cal L}_{ab}$ are given by 
\begin{eqnarray}
  \label{eq:second-order-gauge-inv-metrc-pert-components}
  {\cal L}_{ab}
  &=& 
  - 2 a^{2} \stackrel{(2)}{\Phi} (d\eta)_{a}(d\eta)_{b}
  + 2 a^{2} \stackrel{(2)}{\nu_{i}} (d\eta)_{(a}(dx^{i})_{b)}
  \nonumber\\
  && \quad
  + a^{2} 
  \left( - 2 \stackrel{(2)}{\Psi} \gamma_{ij} 
    + \stackrel{(2)}{{\chi}_{ij}} \right)
  (dx^{i})_{a}(dx^{j})_{b},
\end{eqnarray}
where
$D^{i}\stackrel{(2)}{\nu_{i}}=\stackrel{(2)}{\chi_{[ij]}}=\stackrel{(2)}{\chi^{i}_{\;\;i}}=D^{i}\stackrel{(2)}{\chi_{ij}}
= 0$. 
The vector field $Y_{a}$ is transformed as 
${}_{\;{\cal Y}}\!Y_{a}-{}_{\;{\cal X}}\!Y_{a} = \sigma_{a}$
under the gauge transformations (\ref{eq:Bruni-47-one}) and
(\ref{eq:Bruni-49-one}).


Further, by using the above variables $X_{a}$ and $Y_{a}$, we
can find the gauge invariant variables for the perturbations of
an arbitrary field as 
\begin{eqnarray}
  \label{eq:matter-gauge-inv-def-1.0}
  {}^{(1)}\!{\cal Q} &:=& {}^{(1)}\!Q - {\pounds}_{X}Q_{0},
  , \\ 
  \label{eq:matter-gauge-inv-def-2.0}
  {}^{(2)}\!{\cal Q} &:=& {}^{(2)}\!Q - 2 {\pounds}_{X} {}^{(1)}Q 
  - \left\{ {\pounds}_{Y} - {\pounds}_{X}^{2} \right\} Q_{0}.
\end{eqnarray}
Through the gauge transformation rules (\ref{eq:Bruni-47-one})
and (\ref{eq:Bruni-49-one}), we can easily check that
these variables are gauge invariant up to the first and the
second order, respectively.


As the matter contents, in this letter, we consider the perfect
fluid whose energy-momentum tensor is given by 
\begin{eqnarray}
 \bar{T}_{a}^{\;\;b}
 =
 \left(\bar{\epsilon}+\bar{p}\right) \bar{u}_{a}\bar{u}^{b} 
 + \bar{p}\delta_{a}^{\;\;b}.
 \label{eq:fluid-energy-momentum-tensor-full}
\end{eqnarray}
We expand these fluid components $\bar{\epsilon}$, $\bar{p}$,
and $\bar{u}_{a}$ as Eq.~(\ref{eq:Taylor-expansion-of-calX}):
\begin{eqnarray}
  \bar{\epsilon}
  &=&
  \epsilon
  + \lambda \stackrel{(1)}{\epsilon}
  + \frac{1}{2} \lambda^{2} \stackrel{(2)}{\epsilon} 
  + O(\lambda^{3}),
  \label{eq:Taylor-expansion-of-energy-density}
  \\
  \bar{p}
  &=&
  p
  + \lambda \stackrel{(1)}{p}
  + \frac{1}{2} \lambda^{2} \stackrel{(2)}{p} 
  + O(\lambda^{3})
  ,
  \label{eq:Taylor-expansion-of-pressure}
  \\
  \bar{u}_{a}
  &=&
  u_{a}
  + \lambda \stackrel{(1)}{u}_{a}
  + \frac{1}{2} \lambda^{2} \stackrel{(2)}{u}_{a} 
  + O(\lambda^{3}).
  \label{eq:Taylor-expansion-of-four-velocity}
\end{eqnarray}
Following the definitions (\ref{eq:matter-gauge-inv-def-1.0}) and
(\ref{eq:matter-gauge-inv-def-2.0}), we easily obtain the
corresponding gauge invariant variables for these perturbations
of the fluid components:
\begin{eqnarray}
  \label{eq:kouchan-016.13}
  \stackrel{(1)}{{\cal E}} 
  &:=& \stackrel{(1)}{\epsilon} - {\pounds}_{X}\epsilon, \quad
  \label{eq:kouchan-016.14}
  \stackrel{(1)}{{\cal P}}
  := \stackrel{(1)}{p} - {\pounds}_{X}p, \quad
  \label{eq:kouchan-016.15}
  \stackrel{(1)}{{\cal U}_{a}}
  := \stackrel{(1)}{(u_{a})} - {\pounds}_{X}u_{a}, \nonumber
  \\
  \label{eq:kouchan-016.16}
  \stackrel{(2)}{{\cal E}} 
  &:=& \stackrel{(2)}{\epsilon} 
  - 2 {\pounds}_{X} \stackrel{(1)}{\epsilon}
  - \left\{
    {\pounds}_{Y}
    -{\pounds}_{X}^{2}
  \right\} \epsilon
  , \nonumber
  \\
  \label{eq:kouchan-016.17}
  \stackrel{(2)}{{\cal P}}
  &:=& \stackrel{(2)}{p}
  - 2 {\pounds}_{X} \stackrel{(1)}{p}
  - \left\{
    {\pounds}_{Y}
    -{\pounds}_{X}^{2}
  \right\} p
  , 
  \\
  \label{eq:kouchan-016.18}
  \stackrel{(2)}{{\cal U}_{a}}
  &:=& \stackrel{(2)}{(u_{a})}
  - 2 {\pounds}_{X} \stackrel{(1)}{u_{a}}
  - \left\{
    {\pounds}_{Y}
    -{\pounds}_{X}^{2}
  \right\} u_{a}.
  \nonumber
\end{eqnarray}
Through $\bar{g}^{ab}\bar{u}_{a}\bar{u}_{b}=g^{ab}u_{a}u_{b}=-1$
and neglecting the rotational part in
$\stackrel{(1)}{{\cal U}_{a}}$, we see that 
\begin{equation}
  \stackrel{(1)}{{\cal U}_{a}} = - a \stackrel{(1)}{\Phi} (d\eta)_{a}
  + a D_{i} \stackrel{(1)}{v} (dx^{i})_{a}. 
  \label{eq:first-order-four-velocity-components}
\end{equation}


We also expand the Einstein tensor as 
\begin{eqnarray}
  \bar{G}_{a}^{\;\;b}
  =
  G_{a}^{\;\;b}
  + \lambda {}^{(1)}\!G_{a}^{\;\;b}
  + \frac{1}{2} \lambda^{2} {}^{(2)}\!G_{a}^{\;\;b}
  + O(\lambda^{3}).
\end{eqnarray}
Equations (\ref{eq:linear-metric-decomp}) and
(\ref{eq:H-ab-in-gauge-X-def-second-1}) give each order
perturbation of the Einstein tensor which is decomposed as 
\begin{eqnarray}
  {}^{(1)}\!G_{a}^{\;\;b}
  &=&
  {}^{(1)}{\cal G}_{a}^{\;\;b}\left[{\cal H}\right]
  + {\pounds}_{X}G_{a}^{\;\;b}
  ,
  \\
  {}^{(2)}\!G_{a}^{\;\;b}
  &=&
  {}^{(1)}{\cal G}_{a}^{\;\;b}\left[{\cal L}\right]
  + {}^{(2)}{\cal G}_{a}^{\;\;b}\left[{\cal H}, {\cal H}\right]
  \nonumber\\
  && \quad
  + 2 {\pounds}_{X} {}^{(1)}\!G_{a}^{\;\;b}
  + \left\{
    {\pounds}_{Y}
    -{\pounds}_{X}^{2}
  \right\} G_{a}^{\;\;b}
\end{eqnarray}
as expected from Eqs.~(\ref{eq:matter-gauge-inv-def-1.0}) and
(\ref{eq:matter-gauge-inv-def-2.0}). 
Here, ${}^{(1)}{\cal G}_{a}^{\;\;b}\left[{\cal H}\right]$ and 
${}^{(1)}{\cal G}_{a}^{\;\;b}\left[{\cal L}\right]
+ {}^{(2)}{\cal G}_{a}^{\;\;b}\left[{\cal H}, {\cal H}\right]$
are gauge invariant parts of the frist and the second order
perturbations of the Einstein tensor, respectively.
On the other hand, the energy mometum tensor
(\ref{eq:fluid-energy-momentum-tensor-full}) is also expanded as 
\begin{eqnarray}
  \bar{T}_{a}^{\;\;b}
  =
  T_{a}^{\;\;b}
  + \lambda {}^{(1)}\!T_{a}^{\;\;b}
  + \frac{1}{2} \lambda^{2} {}^{(2)}\!T_{a}^{\;\;b}
  + O(\lambda^{3})
\end{eqnarray}
and ${}^{(1)}\!T_{a}^{\;\;b}$ and ${}^{(2)}\!T_{a}^{\;\;b}$ are
also given in the form 
\begin{eqnarray}
  {}^{(1)}\!T_{a}^{\;\;b}
  &=&
  {}^{(1)}\!{\cal T}_{a}^{\;\;b}
  + {\pounds}_{X}T_{a}^{\;\;b}
  ,
  \\
  {}^{(2)}\!T_{a}^{\;\;b}
  &=&
  {}^{(2)}\!{\cal T}_{a}^{\;\;b}
  + 2 {\pounds}_{X} {}^{(1)}\!T_{a}^{\;\;b}
  + \left\{
    {\pounds}_{Y}
    -{\pounds}_{X}^{2}
  \right\} T_{a}^{\;\;b}
\end{eqnarray}
through the definitions
(\ref{eq:kouchan-016.13})--(\ref{eq:kouchan-016.18}) of the
gauge invariant variables of the fluid components.
Here, ${}^{(1)}\!{\cal T}_{a}^{\;\;b}$ and 
${}^{(2)}\!{\cal T}_{a}^{\;\;b}$ are gauge invariant part of the  
first and the second order perturbation of the energy momentum
tensor, respectively. 
Then, the first and the second order perturbations of the
Einstein equation are necessarily given in term of gauge
invaraint variables:
\begin{eqnarray}
  \label{eq:linear-order-Einstein-equation}
  &&
  {}^{(1)}{\cal G}_{a}^{\;\;b}\left[{\cal H}\right]
  =
  8\pi G {}^{(1)}{\cal T}_{a}^{\;\;b},
  \\
  \label{eq:second-order-Einstein-equation}
  && 
  {}^{(1)}{\cal G}_{a}^{\;\;b}\left[{\cal L}\right]
  + {}^{(2)}{\cal G}_{a}^{\;\;b}\left[{\cal H}, {\cal H}\right]
  =
  8\pi G \;\; {}^{(2)}{\cal T}_{a}^{\;\;b}. 
\end{eqnarray}


The traceless scalar part of the spatial component of
Eq.(\ref{eq:linear-order-Einstein-equation}) yields
$\stackrel{(1)}{\Psi} = \stackrel{(1)}{\Phi}$, and the other
components of Eq.~(\ref{eq:linear-order-Einstein-equation})
gives well-known equations\cite{Bardeen-1980}.
Neglecting the first order vector and tensor modes and using
$\stackrel{(1)}{\Psi} = \stackrel{(1)}{\Phi}$, the components of 
$\stackrel{(2)}{{\cal U}_{a}}$ are given by 
\begin{eqnarray}
  \stackrel{(2)}{{\cal U}_{a}}
  &=&
  a \left(
    \left(\stackrel{(1)}{\Phi}\right)^{2}
    - D_{i}\stackrel{(1)}{v} D^{i}\stackrel{(1)}{v}
    - \stackrel{(2)}{\Phi}
  \right) (d\eta)_{a}
  \nonumber\\
  &&
  +
  a \left(
    D_{i} \stackrel{(2)}{v}
    +
    \stackrel{(2)}{{\cal V}_{i}}
  \right) (dx^{i})_{a},
\end{eqnarray}
where $D^{i} \stackrel{(2)}{{\cal V}_{i}} = 0$.


All components of Eq.~(\ref{eq:second-order-Einstein-equation})
are summarized as follows: 
As the scalar parts of
Eq.~(\ref{eq:second-order-Einstein-equation}), we have 
\begin{widetext}
\begin{eqnarray}
  4\pi G a^{2} \stackrel{(2)}{{\cal E}}
  &=&
  \left(
    - 3 {\cal H} \partial_{\eta}
    +   \Delta
    + 3 K
    - 3 {\cal H}^{2}
  \right)
  \stackrel{(2)}{\Phi}
  - \Gamma_{0}
  +
  \frac{3}{2}
  \left(
    \Delta^{-1} D^{i}D_{j}\Gamma_{i}^{\;\;j}
    - \frac{1}{3} \Gamma_{k}^{\;\;k}
  \right)
  \nonumber\\
  && 
  -
  \frac{9}{2}
  {\cal H} \partial_{\eta}
  \left( \Delta + 3 K \right)^{-1}
  \left(
    \Delta^{-1} D^{i}D_{j}\Gamma_{i}^{\;\;j}
    - \frac{1}{3} \Gamma_{k}^{\;\;k}
  \right)
  \label{eq:kouchan-18.79}
  , \\
  8\pi G a^{2} (\epsilon + p) D_{i}\stackrel{(2)}{v} 
  &=&
  - 2 \partial_{\eta}D_{i}\stackrel{(2)}{\Phi}
  - 2 {\cal H} D_{i}\stackrel{(2)}{\Phi}
  + D_{i} \Delta^{-1} D^{k}\Gamma_{k}
  - 3 \partial_{\eta}D_{i}
  \left( \Delta + 3 K \right)^{-1}
  \left(
    \Delta^{-1} D^{i}D_{j}\Gamma_{i}^{\;\;j}
    - \frac{1}{3} \Gamma_{k}^{\;\;k}
  \right)
  \label{eq:second-velocity-scalar-part-Einstein}
  , \\
  4 \pi G a^{2} \stackrel{(2)}{{\cal P}}
  &=&
  \left(
      \partial_{\eta}^{2} 
    + 3{\cal H} \partial_{\eta}
    - K
    + 2\partial_{\eta}{\cal H}
    + {\cal H}^{2}
  \right)
  \stackrel{(2)}{\Phi}
  -
  \frac{1}{2}
  \Delta^{-1} D^{i}D_{j}\Gamma_{i}^{\;\;j}
  \nonumber\\
  &&
  +
  \frac{3}{2}
  \left(
        \partial_{\eta}^{2} 
    + 2 {\cal H} \partial_{\eta}
  \right)
  \left( \Delta + 3 K \right)^{-1}
  \left(
    \Delta^{-1} D^{i}D_{j}\Gamma_{i}^{\;\;j}
    - \frac{1}{3} \Gamma_{k}^{\;\;k}
  \right)
  ,\\
  \label{eq:kouchan-18.80}
  \stackrel{(2)}{\Psi} - \stackrel{(2)}{\Phi}
  &=&
  \frac{3}{2}
  \left( \Delta + 3 K \right)^{-1}
  \left(
    \Delta^{-1} D^{i}D_{j}\Gamma_{i}^{\;\;j}
    - \frac{1}{3} \Gamma_{k}^{\;\;k}
  \right)
  .
  \label{eq:kouchan-18.65}
\end{eqnarray}
where
\begin{eqnarray}
  \Gamma_{0}
  &:=&
  8\pi G a^{2} (\epsilon + p) D^{i}\stackrel{(1)}{v} D_{i}\stackrel{(1)}{v} 
  - 3 D_{k}\stackrel{(1)}{\Phi} D^{k}\stackrel{(1)}{\Phi}
  - 3 \left(\partial_{\eta}\stackrel{(1)}{\Phi}\right)^{2}
  -  8 \stackrel{(1)}{\Phi} \Delta \stackrel{(1)}{\Phi}
  - 12 \left( K + {\cal H}^{2} \right) \left(\stackrel{(1)}{\Phi}\right)^{2}
  ,
  \label{eq:kouchan-18.54}
  \\
  \Gamma_{i}
  &:=&
  - 16 \pi G a^{2} \left(
     \stackrel{(1)}{{\cal E}} + \stackrel{(1)}{{\cal P}}
   \right)D_{i}\stackrel{(1)}{v}
  +         12  {\cal H} \stackrel{(1)}{\Phi} D_{i}\stackrel{(1)}{\Phi}
  -          4  \stackrel{(1)}{\Phi} \partial_{\eta}D_{i}\stackrel{(1)}{\Phi}
  -          4  \partial_{\eta}\stackrel{(1)}{\Phi} D_{i}\stackrel{(1)}{\Phi}
  ,
  \label{eq:kouchan-18.56}
  \\
  \Gamma_{i}^{\;\;j}
  &:=&
    16 \pi G a^{2} (\epsilon+p) D_{i}\stackrel{(1)}{v} D^{j}\stackrel{(1)}{v}
  -  4  D_{i}\stackrel{(1)}{\Phi} D^{j}\stackrel{(1)}{\Phi}
  -  8  \stackrel{(1)}{\Phi} D_{i}D^{j}\stackrel{(1)}{\Phi}
  \nonumber\\
  && 
  + 2 \left(
      3 D_{k}\stackrel{(1)}{\Phi} D^{k}\stackrel{(1)}{\Phi}
    + 4 \stackrel{(1)}{\Phi} \Delta \stackrel{(1)}{\Phi}
    +   \left(\partial_{\eta}\stackrel{(1)}{\Phi}\right)^{2}
    + 4 \left(
      2 \partial_{\eta}{\cal H} + K + {\cal H}^{2}
    \right) \left(\stackrel{(1)}{\Phi}\right)^{2}
    + 8 {\cal H} \stackrel{(1)}{\Phi} \partial_{\eta}\stackrel{(1)}{\Phi}
  \right)
  \gamma_{i}^{\;\;j}
    .
  \label{eq:kouchan-18.58}
\end{eqnarray}
As the vector parts of
Eq.~(\ref{eq:second-order-Einstein-equation}), we have
\begin{eqnarray}
  8\pi G a^{2} (\epsilon + p) \stackrel{(2)}{{\cal V}}_{i}
  &=&
  \frac{1}{2} \left(
    \Delta
    + 2 K
  \right)
  \stackrel{(2)}{\nu_{i}}
  +\left(
    \Gamma_{i}
    -
    D_{i} \Delta^{-1} D^{k} \Gamma_{k}
  \right)
  \label{eq:kouchan-18.59-vorticity}
  , \\
  \partial_{\eta} \left(a^{2}\stackrel{(2)}{\nu_{i}}\right)
  &=&
  2a^{2}
  \left( \Delta + 2 K \right)^{-1}
  \left\{
    D_{i} \Delta^{-1} D^{k}D_{l}\Gamma_{k}^{\;\;l}
    - D_{k}\Gamma_{i}^{\;\;k}
  \right\}
  .
  \label{eq:kouchan-18.66}
\end{eqnarray}
As the tensor parts of
Eq.~(\ref{eq:second-order-Einstein-equation}), we have the
evolution equation of $\stackrel{(2)\;\;\;\;}{\chi_{ij}}$ 
\begin{eqnarray}
  \left(
    \partial_{\eta}^{2} + 2 {\cal H} \partial_{\eta} + 2 K  - \Delta
  \right)
  \stackrel{(2)\;\;\;\;}{\chi_{ij}}
  &=&
  2 \Gamma_{ij}
  - \frac{2}{3} \gamma_{ij} \Gamma_{k}^{\;\;k}
  - 3
  \left(
    D_{i}D_{j} - \frac{1}{3} \gamma_{ij} \Delta
  \right) 
  \left( \Delta + 3 K \right)^{-1}
  \left(
    \Delta^{-1} D^{k}D_{l}\Gamma_{k}^{\;\;l}
    - \frac{1}{3} \Gamma_{k}^{\;\;k}
  \right)
  \nonumber\\
  &&
  + 4
  \left( 
    D_{(i}\left( \Delta + 2 K \right)^{-1}
    D_{j)}\Delta^{-1}D^{l}D_{k}\Gamma_{l}^{\;\;k}
    - D_{(i} \left( \Delta + 2 K \right)^{-1} D^{k}\Gamma_{j)k}
  \right)
  ,
  \label{eq:kouchan-18.68}
\end{eqnarray}
Equations (\ref{eq:kouchan-18.66}) and (\ref{eq:kouchan-18.68})
imply that the second order vector and tensor modes may be
generated due to the scalar-scalar mode coupling of the first
order perturbation.


Further, the equations (\ref{eq:kouchan-18.79}) and
(\ref{eq:kouchan-18.80}) are reduced to the single equation for
$\stackrel{(2)}{\Phi}$ 
\begin{eqnarray}
  && 
  \left(
      \partial_{\eta}^{2} 
    + 3 {\cal H} (1 + c_{s}^{2}) \partial_{\eta}
    -   c_{s}^{2} \Delta
    + 2 \partial_{\eta}{\cal H}
    + (1 + 3 c_{s}^{2}) ({\cal H}^{2} - K)
  \right)
  \stackrel{(2)}{\Phi}
  \nonumber\\
  &=&
  4\pi G a^{2} \left\{
    \tau \stackrel{(2)}{{\cal S}}
    + \frac{\partial c_{s}^{2}}{\partial\epsilon}
    \left(\stackrel{(1)}{{\cal E}}\right)^{2}
    + 2 \frac{\partial c_{s}^{2}}{\partial S}
    \stackrel{(1)}{{\cal E}}
    \stackrel{(1)}{{\cal S}}
    + \frac{\partial\tau}{\partial S}
    \left(\stackrel{(1)}{{\cal S}}\right)^{2}
  \right\}
  - c_{s}^{2} \Gamma_{0}
  + \frac{1}{6} \Gamma_{k}^{\;\;k}
  +
  \frac{3}{2}
  \left(c_{s}^{2} + \frac{1}{3}\right)
  \left(
    \Delta^{-1} D^{i}D_{j}\Gamma_{i}^{\;\;j}
    - \frac{1}{3} \Gamma_{k}^{\;\;k}
  \right)
  \nonumber\\
  && \quad
  -
  \frac{3}{2}
  \left(
      \partial_{\eta}^{2} 
    + \left( 2 + 3 c_{s}^{2} \right) {\cal H} \partial_{\eta}
  \right)
  \left( \Delta + 3 K \right)^{-1}
  \left(
    \Delta^{-1} D^{i}D_{j}\Gamma_{i}^{\;\;j}
    - \frac{1}{3} \Gamma_{k}^{\;\;k}
  \right)
  .
  \label{eq:second-order-Einstein-scalar-master-eq}
\end{eqnarray}
\end{widetext}
Here, we have used the second order perturbation of the equation
of state for the fluid components
\begin{eqnarray}
  \stackrel{(2)}{{\cal P}}
  =
  c_{s}^{2}
  \stackrel{(2)}{{\cal E}} 
  + 
  \tau
  \stackrel{(2)}{{\cal S}} 
  +
  \frac{\partial c_{s}^{2}}{\partial\epsilon}
  \stackrel{(1)}{{\cal E}}^{2}
  + 2
  \frac{\partial c_{s}^{2}}{\partial S}
  \stackrel{(1)}{{\cal E}}
  \stackrel{(1)}{{\cal S}}
  +
  \frac{\partial\tau}{\partial S}
  \stackrel{(1)}{{\cal S}}^{2},
  \label{eq:second-order-equation-of-state-gauge-inv}
\end{eqnarray}
where $\stackrel{(1)}{{\cal S}}$ and $\stackrel{(2)}{{\cal S}}$
are the gauge invariant entropy perturbation of the first and
second order, respectively, we denoted that $c_{s}^{2} :=
\partial\bar{p}/\partial\bar{\epsilon}$ and $\tau :=
\partial\bar{p}/\partial\bar{S}$.
The equation (\ref{eq:second-order-Einstein-scalar-master-eq})
will be useful to discus the second order effect in the CMB
anisotropy, for example, the non-Gaussianity generated by the
non-linear effects\cite{Non-Gaussianity-in-CMB}.


We gave all components of the second order perturbation of the
Einstein equation in terms of gauge invariant variables.
This is the main result of this paper.
The similar equations are also obtained in the case where the
matter content of the universe is a single scalar
field\cite{kouchan-cosmology-second}.
We also note that we have derived the second order perturbation
equations without using explicit solutions to the first order
perturbations.
Therefore, the equations
(\ref{eq:kouchan-18.79})-(\ref{eq:kouchan-18.68}) are applicable
to any expansion law of the background universe.
In particular, in the case where $a(\eta)=1$, these equations
are those for the second post-Minkowski perturbations.
Besides the explicit forms of the second order Einstein
equations, the gauge invariant expressions 
(\ref{eq:linear-order-Einstein-equation}) and
(\ref{eq:second-order-Einstein-equation}) of the first and the
second order Einstein equation are true if the decomposition
formula (\ref{eq:linear-metric-decomp}) is true.
Hence, the gauge invariant formulation of the second order
perturbation shown here is also applicable not only to
cosmological perturbations but also any other situations of the
general relativistic perturbations.


In the case of cosmological perturbations, 
Eqs.~(\ref{eq:kouchan-18.79})-(\ref{eq:kouchan-18.68}) make 
possible to discuss the non-linear effects in the evolution of
the universe.
In many works\cite{Non-Gaussianity-inflation}, the second order
effects have been investigated through the conserved
quantities which corresponds to the Bardeen parameter in the
linear theory.
It will be interesting to clarify these conservation laws by using the
gauge invariant formulation in this paper.


Further, the rotational part of the fluid velocity in
Eqs.~(\ref{eq:kouchan-18.59-vorticity}) of the vector mode is
also important in the early universe because this part of the
fluid velocity is related to the generation of the magnetic
field in the early universe\cite{Matarrese-etal-2005}.
The generation of the tensor mode by
Eq.~(\ref{eq:kouchan-18.68}) is also interesting, since this is
one of the generation process of gravitational waves.
We have already known that the fluctuations of the scalar mode
exist in the early universe from the anisotropy of the
CMB.
Hence, the generation of the vector mode and tensor mode due to
the second order perturbation will give the lower limit of these
modes in the early universe.


Thus, the formualtion shown here have very wide applications and
we leave these our future works.
These issues are interesting not only from the theoretical point
of veiw but also from the observational point of view. 
Due to its very wide applications, the gauge invariant
formulation shown here will play a key role in general
relativistic perturbations.


\section*{Acknowledgements}
The author acknowledges participants of the workshop on ``Black
Hole, spacetime singularities, cosmic censorship'', which was
held at TIFR, for valuable discussions, in particular, for
Prof. P.S.~Joshi for hospitality during this workshop.
He also thanks to Prof. N.~Dadhich the hospitality during his
visit to IUCAA.
The author deeply thanks to members of DTA at NAOJ and his
family for their encouragement.



%
\end{document}